\def\papertitle{Low-Cost Detection of Degraded Voice Clones via Source--Output Acoustic Consistency}
\def\paperauthorA{Jana Shokr}
\def\paperauthorB{Minos Papadopoulos}
\def\paperauthorC{Jeremy Cooperstock}
\def\paperauthorD{Pavo Orepi\'c}

\documentclass[twoside,a4paper]{article}
\usepackage{etoolbox}

\usepackage{style}
\usepackage[table]{xcolor}
\usepackage{amsmath,amssymb,amsfonts,amsthm}
\usepackage{siunitx}
\usepackage{euscript}
\usepackage[T1]{fontenc}
\usepackage[utf8]{inputenc}
\usepackage[english]{babel}
\usepackage{caption}
\usepackage{subfig}
\usepackage{color}
\usepackage{booktabs}
\usepackage{lipsum}
\usepackage{ifpdf}
\usepackage{times}
\usepackage{dblfloatfix}

\input glyphtounicode
\ninept

\newcounter{numauth}
\newcounter{listcnt}
\newcommand\authcnt[1]{\ifdefined#1 \stepcounter{numauth} \fi}
\newcommand\addauth[1]{
\ifdefined#1
\stepcounter{listcnt}
\ifnum \value{listcnt}<\value{numauth}
\appto\authorslist{, #1}
\else
\appto\authorslist{~and~#1}
\fi
\fi}

\authcnt{\paperauthorB}
\authcnt{\paperauthorC}
\authcnt{\paperauthorD}
\authcnt{\paperauthorE}
\authcnt{\paperauthorF}
\authcnt{\paperauthorG}
\authcnt{\paperauthorH}
\authcnt{\paperauthorI}
\authcnt{\paperauthorJ}

\def\authorslist{\paperauthorA}
\addauth{\paperauthorB}
\addauth{\paperauthorC}
\addauth{\paperauthorD}
\addauth{\paperauthorE}
\addauth{\paperauthorF}
\addauth{\paperauthorG}
\addauth{\paperauthorH}
\addauth{\paperauthorI}
\addauth{\paperauthorJ}

\newif\ifpdf
\ifx\pdfoutput\relax
\else
   \ifcase\pdfoutput
      \pdffalse
   \else
      \pdftrue
   \fi
\fi

\ifpdf
  \usepackage[pdftex,
    pdftitle={\papertitle},
    pdfauthor={\authorslist},
    pdfsubject={arXiv preprint},
    colorlinks=true,
    citecolor=red,
    linkcolor=red,
    urlcolor=red,
    bookmarksnumbered,
    pdfstartview=XYZ
  ]{hyperref}
  \usepackage[pdftex]{graphicx}
\else
  \usepackage[dvips]{epsfig,graphicx}
  \usepackage[dvips,
    pdftitle={\papertitle},
    pdfauthor={\authorslist},
    pdfsubject={arXiv preprint},
    colorlinks=true,
    citecolor=red,
    linkcolor=red,
    urlcolor=red,
    bookmarksnumbered,
    pdfstartview=XYZ
  ]{hyperref}
\fi

\usepackage[hypcap=true]{caption}
\title{\papertitle}

\affiliation
{Jana Shokr$^{1}$, Minos Papadopoulos$^{1}$, Jeremy Cooperstock$^{1}$, Pavo Orepi\'c$^{2,*}$\thanks{Corresponding author: \href{mailto:pavo.orepic@uzh.ch}{pavo.orepic@uzh.ch}. This manuscript has not been peer reviewed and has not been accepted for publication.}}
{$^{1}$Department of Electrical and Computer Engineering, McGill University, Montreal, Canada\\
$^{2}$Linguistic Research Infrastructure (LiRI), University of Zurich, Zurich, Switzerland}

\begin{document}

\ifpdf
  \DeclareGraphicsExtensions{.png,.jpg,.pdf}
\else
  \DeclareGraphicsExtensions{.eps}
\fi

\maketitle
\begin{abstract}Recent advances in generative speech have increased the need for automatic detection of obviously failed synthetic outputs. This is particularly important in clinical settings such as AVATAR therapy, in which schizophrenia patients engage with a computer-generated representation of their hallucinated voices and degraded synthesis may disrupt immersion and therapeutic engagement. We investigate whether low-dimensional, interpretable source--output acoustic features can provide a lightweight first-pass detector of degraded voice-cloning outputs.
Motivated by source--filter models of speech, we first test median fundamental frequency ($f_0$) as a source-related consistency measure, and compare it with vocal tract length (VTL) as a filter-related measure and Harmonics-to-Noise Ratio (HNR) as a noise-related descriptor. Human-labeled voice-cloning samples generated with two vocoder families, WaveRNN ($n=54$) and HiFi-GAN ($n=40$), were evaluated using an asymmetric thresholding procedure in the input--output feature space.
For WaveRNN, $f_0$ and HNR both achieved 85.2\% accuracy, outperforming VTL (64.8\%). For HiFi-GAN, HNR achieved 80.0\% accuracy, followed by $f_0$ at 77.5\% and VTL at 67.5\%. Sample-level overlap and spectrographic inspection showed that $f_0$ and HNR capture partly distinct failure patterns, rather than providing redundant rankings of the same samples.
These results show that simple source--output acoustic consistency measures can provide useful first-pass detection of degraded voice clones, and support the use of interpretable threshold-based screening in applications where failed synthetic speech must be rejected quickly.
\end{abstract}

\section{Introduction}
\label{sec:intro}

Recent advances in generative artificial intelligence have led to the widespread use of synthetic speech in applications such as conversational agents, speech-to-speech translation, audiobook and dubbing production, personalized voice cloning, augmentative and alternative communication, digital avatars, and social robots. The challenge is not only to generate natural speech under ideal conditions, but also to detect outputs that fail in obvious ways. In practice, such failures may sound robotic, unstable, noisy, or otherwise implausible, diminishing intelligibility and undermining user trust. The problem is particularly important in systems that operate at scale or under real-time constraints, where manual listening to every generated sample is not feasible. This creates a need for lightweight and interpretable quality-control methods that can identify clearly failed outputs automatically before they are presented to the user.

This need is especially important in clinical applications of synthetic speech, including personalized voices for augmentative and alternative communication, voice restoration, and therapeutic virtual agents. In such settings, degraded output is not merely inconvenient, but may interfere with communication and engagement, thereby affecting clinical outcomes. A notable example is the AVATAR therapy \cite{Craig2018} for auditory-verbal hallucinations (AVH), colloquially referred to as “hearing voices”. AVH are among the most common and distressing symptoms of psychosis, often resulting in suicide \cite{Harkavy-Friedman2003}. In recent years, AVATAR therapy has emerged as a promising clinical approach in which patients engage in dialogue with a computer-generated representation of their hallucinated voice, allowing therapeutic confrontation in a controlled setting \cite{Craig2018, giguere_reattribution_2025, garety_digital_2024}. For such an interaction to be effective, the synthesized voice must be sufficiently natural and credible to preserve immersion and emotional relevance \cite{huckvale_avatar_2013, Lee2022}. If the signal contains strong artifacts or an obviously artificial quality, the therapeutic encounter may be disrupted. Thus, in clinical contexts, quality control is not merely a matter of technical refinement, but a practical requirement for safe and effective use.

Here, we aimed to develop an efficient method for identifying and discarding clearly degraded vocoder outputs in AVATAR therapy systems \cite{Lee2022},
supporting a more reliable and efficient use. Specifically, we addressed a practical signal-processing problem arising in voice-based therapeutic systems: how to reject clearly failed synthetic samples quickly and automatically before they reach the end user. More broadly, this work asks whether low-dimensional acoustic features grounded in established accounts of speech production and voice perception can provide a lightweight and interpretable first-pass detector of severe voice-cloning failures.

\section{Related Work}

In speech synthesis and voice conversion research, quality and naturalness are still most commonly assessed through subjective listening tests, especially Mean Opinion Score (MOS) \cite{cooper_review_2024}. While recent work has introduced automatic, non-intrusive predictors of perceived speech quality \cite{huang_voicemos_2022, mittag_nisqa_2021}, these methods generally target overall quality estimation rather than the rapid rejection of clearly failed samples. This creates a clear need for an acoustic measure that is computationally lightweight, transparent, and effective at detecting obvious synthesis failures.

A descriptor related directly to noisiness or signal degradation, such as Harmonics-to-Noise Ratio (HNR), would appear to be a natural candidate. However, synthesis failures do not necessarily manifest only as added noise. In source-filter accounts of speech production, the signal reflects the interaction between an excitation source generated by vocal-fold vibration and the resonant filtering imposed by the vocal tract \cite{Ghazanfar2008}. Within this framework, fundamental frequency ($f_0$) is the most direct acoustic correlate of the source and a perceptually salient cue for characterizing voices \cite{baumann_perceptual_2010, kreiman_information_2024}. In particular, perceptual studies of voice identity have shown that low-dimensional voice spaces are organized in large part by source-related and filter-related dimensions, with $f_0$ and formant structure providing major contributions \cite{baumann_perceptual_2010, latinus_norm-based_2013, orepic_bone_2023}. It follows that, if vocoded synthesis remains perceptually faithful to the input voice, broad $f_0$ structure should be preserved even when other aspects of the signal vary.

\section{Present approach}

We therefore asked, as a first step, whether fundamental frequency alone could provide a useful low-cost indicator of vocoder failure. Specifically, we tested whether source--output consistency in median $f_0$ was sufficient to distinguish outputs judged by human listeners as acceptable from those judged as degraded because of audible artifacts or unnaturalness. This hypothesis was evaluated on human-labeled voice-cloning samples generated with two vocoder families, WaveRNN \cite{WaveRNN} and HiFi-GAN \cite{HiFiGAN}, in order to test whether the proposed $f_0$-based criterion remained informative across vocoders with different synthesis architectures and operating characteristics.

The same approach was then applied to vocal tract length (VTL) and HNR features to contextualize the role of $f_0$. Specifically, because source--filter models of speech \cite{Ghazanfar2008, kreiman_information_2024} motivate $f_0$ as a salient source-related cue, VTL was included as a corresponding filter-related measure of vocal-tract characteristics. HNR was included as a more intuitive noise-related descriptor, allowing us to test whether a measure of harmonicity and aperiodicity would provide complementary information to $f_0$ in detecting degraded outputs. Unlike $f_0$, which indexes the periodic rate of vocal-fold vibration and thus the pitch-bearing source structure, HNR reflects the relative strength of harmonic versus noise-like components in the signal and is therefore sensitive to a different class of synthesis artifacts.

\section{Method}

\subsection{Datasets and Vocoder Architectures}
To evaluate the proposed pitch-based evaluation of the synthesis output, a dataset of human-labeled audio samples was utilized. Source utterances were drawn from the LibriSpeech ASR corpus, a large-scale audiobook-based dataset of read English speech, and were used to generate the synthesized samples later labeled by human listeners as “good” or “bad.” \cite{WaveRNN}. To validate the generalizability of our heuristic, a secondary dataset ($n = 40$, split equally between labels) was generated using HiFi-GAN \cite{HiFiGAN}. Unlike sequential models, HiFi-GAN is a non-autoregressive, GAN-based vocoder that utilizes a Multi-Period Discriminator (MPD) and Multi-Receptive Field Fusion (MRF) to synthesize waveforms in parallel \cite{HiFiGAN}.
Since these two vocoders represent fundamentally different generation paradigms, sequential versus parallel, they are expected to produce distinct artifact profiles.
By testing against both serial and parallel architectures, we aimed to determine whether $f_0$ drift can serve as a robust, cross-vocoder heuristic for failure detection. Binary labels (''good'' vs.\ bad'') were assigned by human listeners based on the presence of major audible digital artifacts or ``roboticism.''

\subsection{Automated Acoustic Feature Extraction}
A feature extraction pipeline was developed using Python. To extract fundamental frequency ($f_0$) and vocal-tract length (VTL) features, we used the Parselmouth library \cite{Parselmouth}, which is designed to directly interface with the Praat phonetic software framework \cite{Praat}. The extraction algorithm explicitly filtered out unvoiced segments ($f_0 = 0$ Hz),  ensuring that the median value accurately reflected the sustained pitch of the synthesized speech \cite{Quatieri}.
 Harmonic-to-Noise Ratio (HNR)\cite{harmonicity} was extracted using Praat's  cross-correlation method \texttt{to\_harmonicity\_cc()}, which estimates short-term acoustic periodicity in dB. To obtain a single utterance-level HNR estimate, frame-wise values were averaged across all valid voiced segments. Frames representing silence or undefined periodicity were excluded to ensure the mean reflected the harmonic quality of the speech signal only \cite{harmonicity}.

\subsection{A low-cost
pitch-based evaluation of voice clones}

As noted above, we hypothesized that the fundamental frequency plays an important role in the perception of degraded voices.
Visually contrasting the spectrograms of good and bad voice output samples,
we observed that pitch extraction from the bad samples is problematic for conventional speech analysis tools.
This was borne out by our attempts to extract $f_0$ with Praat.
Curious to explore further, we  extracted the $f_0$ from several badly synthesized samples

and noticed that these values tended to exhibit a significant difference from the $f_0$ of the corresponding input voice sample, including physiologically impossible values and high variability across time. This motivated our further exploration of the lightweight evaluation method we propose here, specifically,
that high-quality synthetic voices preserve the fundamental frequency of the source, while degraded samples exhibit significant differences in $f_0$
between the source input and the synthetic output.

To determine an objective decision boundary, we implemented a two-sided asymmetric thresholding procedure in the input--output feature space. Let $X$ denote the input feature value and $Y$ the corresponding synthesized output value. Perfect preservation lies on the identity line,
\[
Y = X.
\]
Classification was based on the signed deviation from this line,
\[
d = Y - X,
\]
so that $d>0$ indicates samples above the identity line and $d<0$ samples below it.

The classifier was defined by two lines parallel to the identity line,
\[
Y = X + T_{\mathrm{neg}}
\qquad \text{and} \qquad
Y = X + T_{\mathrm{pos}},
\]
with $T_{\mathrm{neg}} < 0$ and $T_{\mathrm{pos}} > 0$. These two thresholds define an acceptance band around $Y=X$. Samples were classified as ``good'' if
\[
T_{\mathrm{neg}} \le d \le T_{\mathrm{pos}},
\]
and as ``bad'' otherwise. Because positive and negative deviations represent distinct directions of mismatch, the lower and upper thresholds were optimized independently: $T_{\mathrm{neg}}$ using only samples with $d<0$, and $T_{\mathrm{pos}}$ using only samples with $d>0$. In each half-space, the threshold yielding the highest classification accuracy against the human labels was retained.

To validate the resulting classifier, the algorithmic classifications were benchmarked against the human labels to allow for the construction of a formal confusion matrix, from which the following diagnostic performance metrics were derived \cite{ROC}: True Positives ($TP$) and True Negatives ($TN$) represent instances where the asymmetric thresholds correctly identified ``bad'' and ``good'' samples, respectively. False Positives ($FP$) and False Negatives ($FN$) represent the model's classification errors relative to human perception. From these values, Overall Accuracy was calculated as $(TP + TN) / N$ to measure the total proportion of correct classifications. To further assess the model's reliability in detecting degradation, Sensitivity was computed as $TP / (TP + FN)$, representing the model's ability to correctly flag ``bad'' samples. Conversely, Specificity was calculated as $TN / (TN + FP)$ to determine the model's accuracy in identifying ``good'' samples.

The same classifier procedure and evaluation metrics were then applied to VTL and HNR features.

\section{Results}
The goal of this analysis was to validate a low-cost, interpretable method for detecting degraded voice-cloning outputs from simple source--output acoustic features. We first report the reliability of the human labels used as ground truth, and then evaluate the threshold-based classifiers for $f_0$, HNR, and VTL across the two vocoder families.

\subsection{Human label agreement}

The binary labeling of these samples was established through an inter-rater agreement process involving four independent listeners. As shown in Table~\ref{table: human_classif}, agreement was high for both vocoders, with especially strong consensus for degraded samples. For WaveRNN, raters agreed on 84.04\% of good samples and 98.15\% of bad samples. For HiFi-GAN, agreement was 88.75\% for good samples and 93.75\% for bad samples. This indicates that the distinction between acceptable and clearly degraded outputs was sufficiently stable to support threshold-based analysis, and also suggests that bad samples were generally easier for listeners to identify consistently than good samples.

\begin{table}[ht]
\centering
\renewcommand{\arraystretch}{1.2}
\caption{Consensus of Human Subjective Quality Labels (4 Raters)}
\vspace{-2mm}
\label{table: human_classif}
\begin{tabular}{|l|c|c|}
\hline
\textbf{Category} & \textbf{Agreement} & \textbf{Acc. (\%)} \\ \hline
\multicolumn{3}{|c|}{ \cellcolor[gray]{0.9}\textbf{WaveRNN (Primary) (n=54)}} \\ \hline
Good Samples      & 88 / 108           & 84.04              \\ \hline
Bad Samples       & 106 / 108          & 98.15              \\ \hline
\multicolumn{3}{|c|}{ \cellcolor[gray]{0.9}\textbf{HiFi-GAN (Validation) (n=40)}} \\ \hline
Good Samples      & 71 / 80            & 88.75              \\ \hline
Bad Samples       & 75 / 80            & 93.75              \\ \hline
\end{tabular}
\end{table}

\subsection{Feature-space distributions and asymmetric thresholds}

\begin{figure*}
    \centering
    \includegraphics[width=0.9\linewidth]{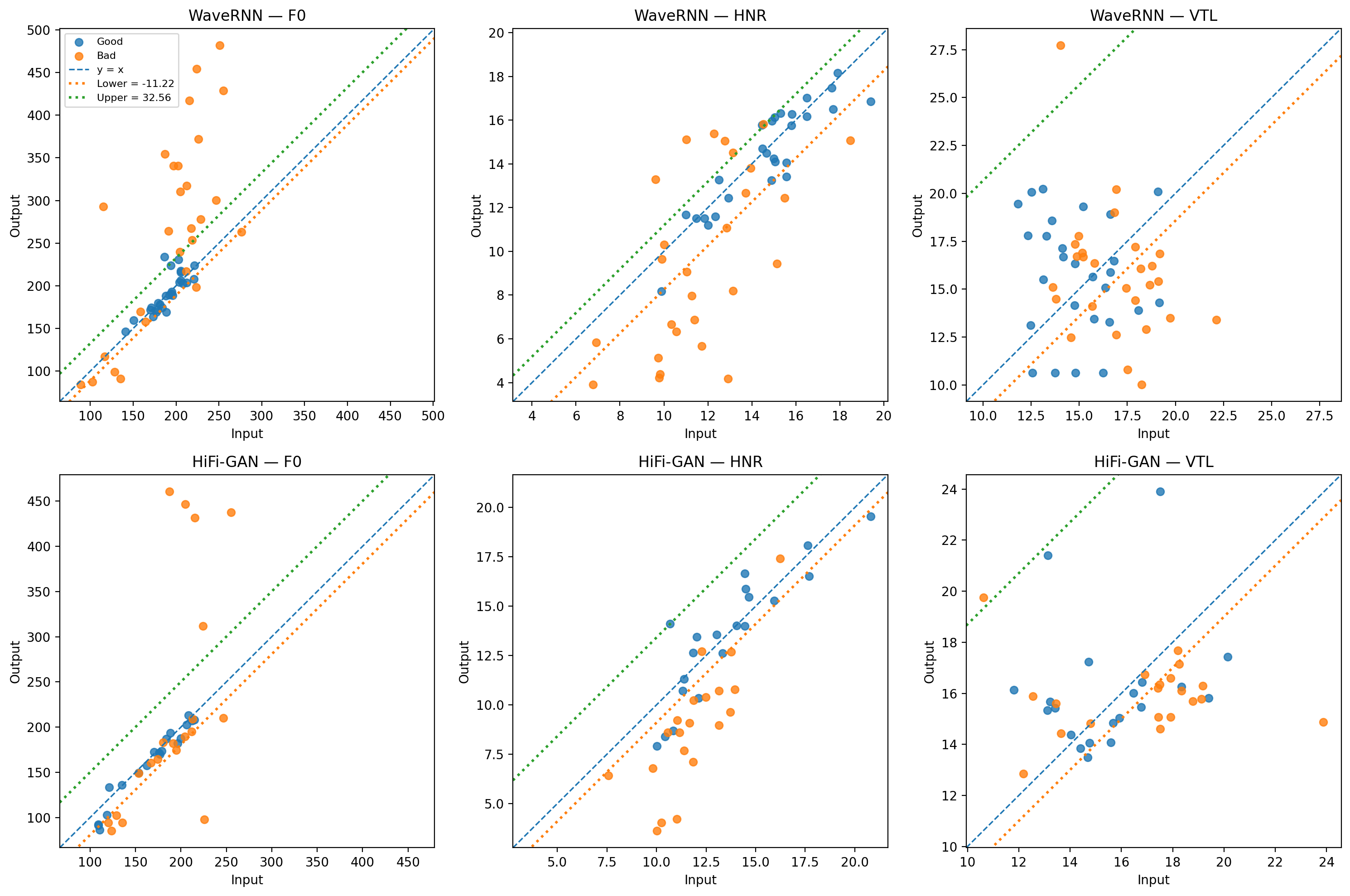}
    \caption{Scatter plots comparing input versus output values for F0, HNR, and VTL parameters across WaveRNN (top row) and HiFi-GAN (bottom row) synthesized voices. Blue and orange markers indicate ``Good'' and ``Bad'' quality samples, respectively. They were calculated to maximize overall classification accuracy by minimizing the sum of false positive and false negative errors relative to the human perceptual ground truth. The dashed blue diagonal represents the identity line ($y = x$), while the dotted orange and green lines indicate lower and upper thresholds for acceptable variation.}
    \label{fig:six-plots}
\end{figure*}

Figure~\ref{fig:six-plots} shows the input--output distributions for the three acoustic features under the independently optimized asymmetric thresholds. Across both vocoders, the clearest geometric structure was observed for $f_0$: samples labeled as good clustered relatively tightly around the identity line, whereas bad samples were more often displaced beyond the acceptance band. This pattern was particularly pronounced for WaveRNN, where many bad samples showed large positive deviations in output pitch.

The optimized thresholds and classification metrics are summarized in Table~\ref{table:full_metrics_analysis}. For WaveRNN, both $f_0$ and HNR achieved the highest overall accuracy, at 85.19\%, whereas VTL performed substantially worse at 64.81\%. For HiFi-GAN, HNR achieved the highest overall accuracy, at 80.00\%, followed closely by $f_0$ at 77.50\%, while VTL again showed weaker performance (67.50\%). Thus, VTL was consistently the least informative feature, whereas $f_0$ and HNR emerged as the two strongest candidates.

\begin{table*}
\centering
\small
\renewcommand{\arraystretch}{1.3}
\caption{Comprehensive Asymmetric Threshold and Confusion Matrix Analysis. Neg T.\ and Pos T.\ refer to negative and positive thresholds, respectively.}
\label{table:full_metrics_analysis}
\begin{tabular}{|l|c|c|c|c|c|c|c|c|c|c|}
\hline
\textbf{Feat.} & \textbf{Neg T.} & \textbf{Pos T.} & \textbf{Acc.} & \textbf{Sens.} & \textbf{Spec.} & \textbf{TP} & \textbf{TN} & \textbf{FP} & \textbf{FN} \\ \hline
\multicolumn{10}{|c|}{\cellcolor[gray]{0.9}\textbf{WaveRNN}} \\ \hline
$f_0$  & -11.2 & 32.6 & 0.85 & 0.82 & 0.89 & 22 & 24 & 3 & 5 \\ \hline
HNR & -1.7  & 1.2  & 0.85 & 0.82 & 0.89 & 22 & 24 & 3 & 5 \\ \hline
VTL & -1.4  & 10.7 & 0.65 & 0.60 & 0.70 & 16 & 19 & 8 & 11 \\ \hline
\multicolumn{10}{|c|}{\cellcolor[gray]{0.9}\textbf{HiFi-GAN}} \\ \hline
$f_0$  & -19.3 & 50.1 & 0.78 & 0.60 & 0.95 & 12 & 19 & 1 & 8 \\ \hline
HNR & -0.9  & 3.4  & 0.80 & 0.90 & 0.70 & 18 & 14 & 6 & 2 \\ \hline
VTL & -1.0  & 8.7  & 0.68 & 0.65 & 0.70 & 13 & 14 & 6 & 7 \\ \hline
\end{tabular}
\end{table*}

The diagnostic metrics revealed an important difference between these two features. For HiFi-GAN, $f_0$ showed higher specificity (95.00\%) than HNR (70.00\%), indicating that pitch consistency was better at retaining acceptable samples and produced fewer false alarms. In contrast, HNR showed higher sensitivity (90.00\%) than $f_0$ (60.00\%), indicating that it captured a larger fraction of degraded outputs. A similar, though less pronounced, trade-off was also visible in the WaveRNN data. Taken together, these results suggest that the two features are not simply interchangeable, but emphasize different aspects of synthesis failure.

\subsection{Complementarity of $f_0$ and HNR}

\begin{figure*}
    \centering
    \includegraphics[width=\textwidth]{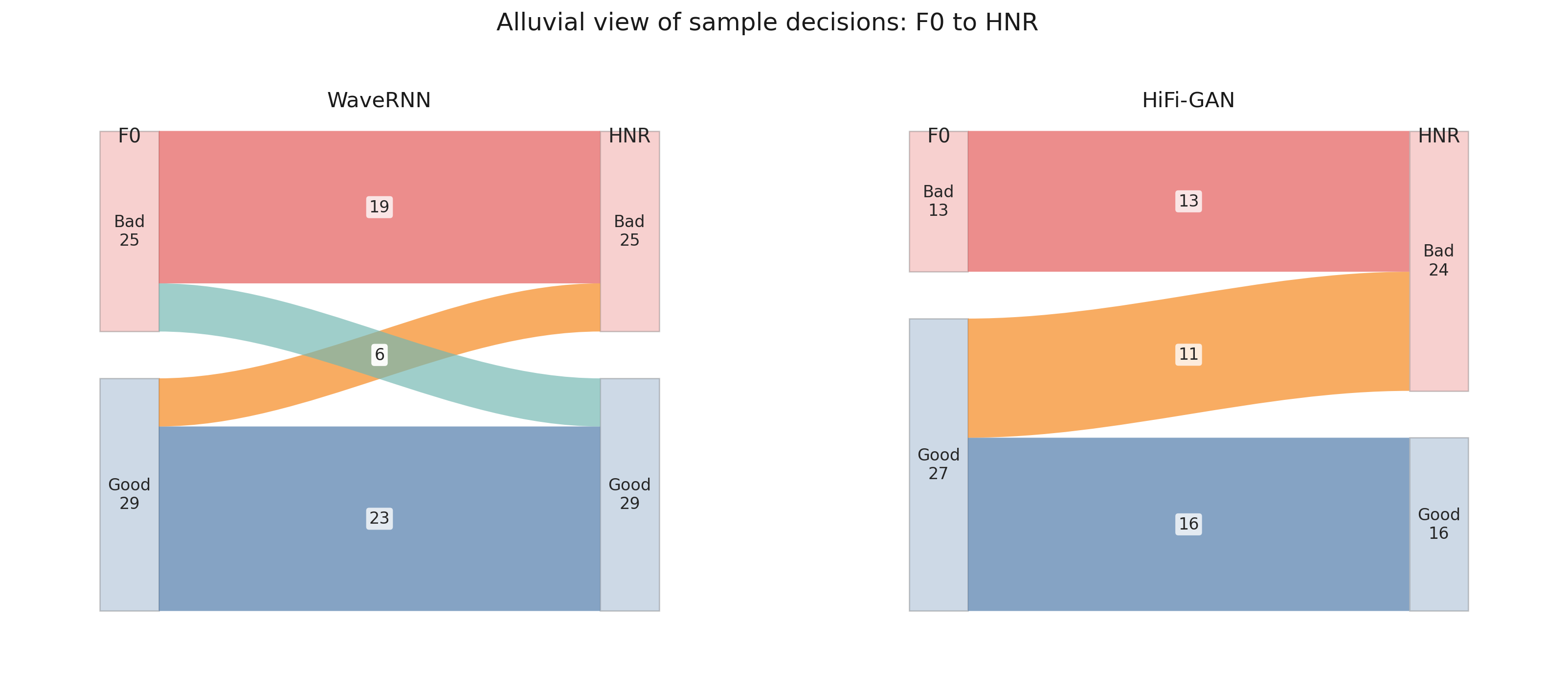}
    \caption{Alluvial view of sample-level agreement between the f0- and HNR-based classifiers. Flows show how samples labeled as ``good'' or ``bad'' by $f_0$ are redistributed under HNR. For WaveRNN, disagreement is approximately balanced in both directions, whereas for HiFi-GAN it is asymmetric, with additional samples rejected by HNR but not by $f_0$.}
    \label{fig:alluvial}
\end{figure*}

To examine this more directly, we compared the sample-level decisions of the $f_0$- and HNR-based classifiers using the alluvial visualization in Figure~\ref{fig:alluvial}. For WaveRNN, disagreement was approximately balanced in both directions: six samples accepted by $f_0$ were rejected by HNR, and six samples rejected by $f_0$ were accepted by HNR. This indicates that the two features captured partly distinct subsets of failures, despite having the same overall accuracy.

For HiFi-GAN, the disagreement pattern was markedly asymmetric. Thirteen samples were rejected by both classifiers and sixteen were accepted by both, but eleven additional samples accepted by $f_0$ were rejected by HNR, whereas no samples rejected by $f_0$ were recovered by HNR. This asymmetry is consistent with the higher sensitivity and lower specificity of HNR in the HiFi-GAN condition, and suggests that HNR responds to degradations that are not necessarily reflected in gross pitch mismatch alone.

These sample-level patterns show that similar overall accuracy can mask qualitatively different classifier behavior. In particular, $f_0$ and HNR do not merely provide redundant views of the same failure cases, but appear to detect different types of mismatch between source and output.

\subsection{Representative failure modes}

\begin{figure*}[h!]
    \centering
    \includegraphics[width=\textwidth]{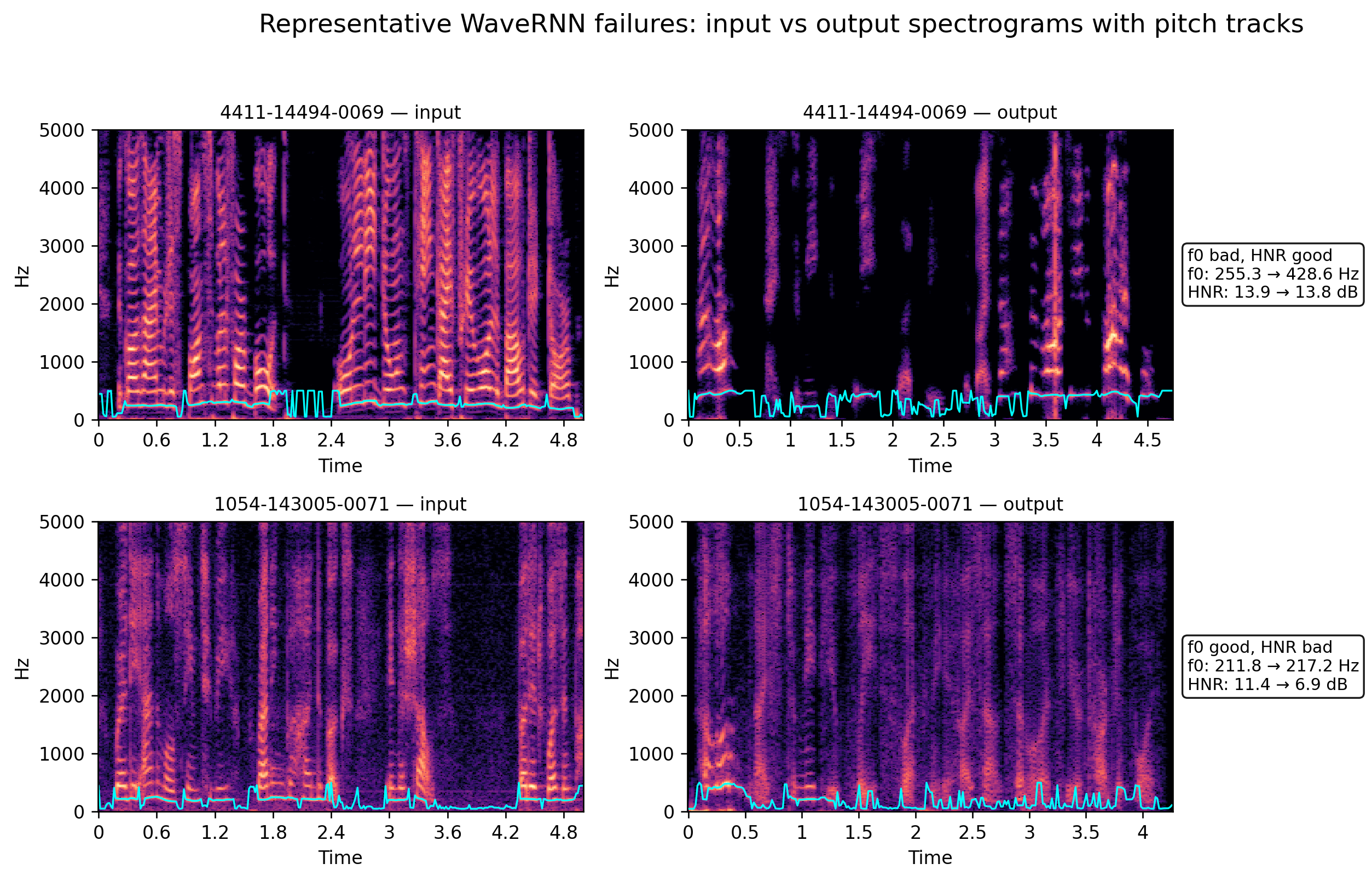}
    \caption{Representative WaveRNN failure modes illustrating the complementary behavior of the two feature-based classifiers. Top row: a sample rejected by the f0-based threshold but accepted by HNR, showing a large source--output pitch mismatch despite retained harmonic structure. Bottom row: a sample accepted by the f0-based threshold but rejected by HNR, showing broadly preserved pitch together with reduced harmonic clarity and more diffuse spectral organization. Cyan curves indicate estimated pitch tracks.}
    \label{fig:spectograms}
\end{figure*}

Figure~\ref{fig:spectograms} illustrates two representative WaveRNN failures that clarify the distinction between the two features. In the top row, the sample is rejected by the $f_0$-based classifier but accepted by HNR. Here, the output retains visible harmonic structure, but the estimated pitch track is shifted strongly upward relative to the input (255.3~Hz to 428.6~Hz), indicating a large source-related mismatch despite preserved harmonicity. In the bottom row, the sample is accepted by the $f_0$-based classifier but rejected by HNR. In this case, the gross pitch remains close to the input (211.8~Hz to 217.2~Hz), but the output spectrogram is more diffuse and exhibits weaker harmonic contrast, consistent with the substantial drop in HNR (11.4~dB to 6.9~dB).

These examples support the interpretation suggested by the quantitative results. The $f_0$-based criterion is particularly sensitive to failures in source preservation, especially large shifts in pitch structure, whereas HNR is more sensitive to reductions in harmonic clarity and increased noise-like or weakly periodic spectral organization. In this sense, the two measures are complementary: $f_0$ acts more strongly as a detector of source-related mismatch, while HNR captures degradations that preserve gross pitch but reduce perceived naturalness.

\section{Discussion}

The present results show that a simple thresholding procedure applied to low-dimensional acoustic features can provide a useful first-pass detector of degraded voice-cloning outputs. Across both vocoders, $f_0$ and HNR were the two most informative features, whereas VTL was consistently less discriminative. This suggests that synthesis failures are reflected more clearly in source-related pitch structure and harmonic integrity than in the vocal-tract-related descriptor tested here.

The performance of $f_0$ is consistent with its role in source--filter accounts of speech production and in perceptual models of voice identity. Because fundamental frequency is a direct acoustic correlate of the excitation source, large source--output deviations in median $f_0$ indicate that the synthesis process has failed to preserve an important aspect of the original voice. In this sense, the proposed $f_0$-based measure is not only computationally lightweight, but also interpretable: it provides a direct indication of whether the generated output remains sufficiently close to the source in one of the most salient dimensions of vocal structure.

At the same time, the comparison with HNR shows that pitch consistency alone does not capture all relevant degradations. The sample-level overlap analysis demonstrated that the $f_0$- and HNR-based classifiers were only partially redundant, and the disagreement patterns differed across vocoders. For WaveRNN, disagreement was approximately balanced in both directions, indicating that the two features captured partly distinct subsets of degraded samples. For HiFi-GAN, by contrast, HNR rejected additional samples that were accepted by $f_0$, suggesting that some failures affected harmonic clarity or periodicity without strongly disrupting gross pitch structure. The spectrographic examples support this interpretation: samples rejected by $f_0$ but accepted by HNR showed large pitch shifts despite retained harmonic structure, whereas samples rejected by HNR but accepted by $f_0$ preserved broad pitch range while exhibiting weaker harmonic contrast and more diffuse spectral organization. Together, these findings indicate that $f_0$ and HNR emphasize different aspects of synthesis failure.

Inspection of representative WaveRNN spectrograms (Figure~\ref{fig:spectograms}) suggests that the two features are not merely correlated alternatives but respond to different structural failures. Samples rejected by $f_0$ but accepted by HNR retained locally harmonic voiced segments, yet exhibited large pitch shifts and intermittent voicing, indicating a breakdown in source preservation rather than harmonicity per se. By contrast, samples rejected by HNR but accepted by $f_0$ preserved the overall pitch range of the input while showing weakened harmonic contrast, smeared spectral structure, or fragmented periodicity, revealing degradations that a median-$f_0$ summary can miss.

The results also suggest that the effective threshold range is vocoder-dependent. The optimized $f_0$ and HNR thresholds differed between WaveRNN and HiFi-GAN, indicating that the acceptance band around the identity line is not fixed across models. One possible explanation is that different vocoder architectures produce different characteristic artifact profiles. The two vocoder architectures differ fundamentally in the way they convert mel spectrograms to audio.
These differences have direct consequences on the types of artifacts produced. WaveRNN is an auto-regressive model built on a gated recurrent unit (GRU), which generates waveform samples one at a time, each conditioned on all preceding samples. In contrast, HiFi-GAN is a non-auto-regressive, fully convolutional generator that synthesizes the entirety of the waveform in one forward pass through a stack of upsampling transposed convolutions interleaved with Multi-Receptive Field Fusion (MRF) residual blocks that operate at multiple kernel sizes and dilation rates.
The main practical implication is that threshold calibration may need to be performed separately for each vocoder family rather than assumed to generalize directly across models.

Beyond the immediate classification results, the present findings are important because they address a practical barrier to the reliable use of synthetic speech in high-stakes settings. A lightweight detector of clearly failed outputs can reduce the need for exhaustive manual screening and make voice-cloning systems more usable in contexts where rapid verification is required. This is especially relevant for AVATAR therapy, where the therapeutic effect depends not only on semantic content but also on the credibility and emotional salience of the generated voice. In such a setting, an output that sounds robotic, unstable, or otherwise implausible is not merely a technical defect: it may weaken immersion, reduce engagement, or disrupt the therapeutic encounter altogether. An interpretable first-pass screening method therefore has potential value not only for improving efficiency, but also for supporting safer and more reliable clinical deployment.

Several limitations of this approach should be noted.

First, the datasets remain relatively small, particularly for HiFi-GAN, and the stability of the optimized thresholds should therefore be confirmed on larger and more balanced sample sets. Second, the present analysis was restricted to a small set of low-dimensional features, and did not compare directly against learned speech-quality predictors. Third, while the spectrographic examples help interpret the disagreement between $f_0$ and HNR, they are illustrative rather than exhaustive. Future work should therefore test the approach on additional vocoders, larger labeled datasets, and combined feature sets in order to determine whether source-related and harmonicity-related measures can be integrated into a more robust multi-feature screening pipeline.

\section{Conclusion}
Overall, our results show that a simple asymmetric thresholding procedure in the input--output feature space can recover human judgments of synthesis quality with useful accuracy. Among the three features, $f_0$ and HNR were consistently the most informative, while VTL was weaker across both vocoders. The results further indicate that $f_0$ and HNR reflect different trade-offs between sensitivity and specificity, and that their partial disagreement is meaningful rather than incidental. This supports the view that low-dimensional acoustic features can provide an effective first-pass detector of vocoder failure, while also suggesting that source-related and harmonicity-related cues capture different aspects of degraded synthetic speech.


\begin{center}
{\bfseries FUNDING}
\end{center}

This project was supported by funding from the Natural Sciences and Engineering Research Council (NSERC) Canada (to J. Cooperstock) and from the Swiss National Science Foundation Grant CRSK-1\_227695 (to P. Orepic).

\bibliographystyle{IEEEtran}
\bibliography{converted, references_pavo}

\begin{thebibliography}{10}
\providecommand{\url}[1]{#1}
\csname url@samestyle\endcsname
\providecommand{\newblock}{\relax}
\providecommand{\bibinfo}[2]{#2}
\providecommand{\BIBentrySTDinterwordspacing}{\spaceskip=0pt\relax}
\providecommand{\BIBentryALTinterwordstretchfactor}{4}
\providecommand{\BIBentryALTinterwordspacing}{\spaceskip=\fontdimen2\font plus
\BIBentryALTinterwordstretchfactor\fontdimen3\font minus \fontdimen4\font\relax}
\providecommand{\BIBforeignlanguage}[2]{{%
\expandafter\ifx\csname l@#1\endcsname\relax
\typeout{** WARNING: IEEEtran.bst: No hyphenation pattern has been}%
\typeout{** loaded for the language `#1'. Using the pattern for}%
\typeout{** the default language instead.}%
\else
\language=\csname l@#1\endcsname
\fi
#2}}
\providecommand{\BIBdecl}{\relax}
\BIBdecl

\bibitem{Craig2018}
\BIBentryALTinterwordspacing
T.~K. Craig, M.~Rus-Calafell, T.~Ward, J.~P. Leff, M.~Huckvale, E.~Howarth, R.~Emsley, and P.~A. Garety, ``{AVATAR} therapy for auditory verbal hallucinations in people with psychosis: a single-blind, randomised controlled trial,'' \emph{The Lancet Psychiatry}, vol.~5, no.~1, pp. 31--40, 2018, publisher: The Author(s). Published by Elsevier Ltd. This is an Open Access article under the CC BY 4.0 license. [Online]. Available: \url{http://dx.doi.org/10.1016/S2215-0366(17)30427-3}
\BIBentrySTDinterwordspacing

\bibitem{Harkavy-Friedman2003}
J.~Harkavy-Friedman, D.~Kimhy, E.~Nelson, D.~Venarde, D.~Malaspina, and J.~Mann, ``Suicide attempts in schizophrenia: the role of command auditory hallucinations for suicide.'' \emph{The Journal of clinical psychiatry.}, vol.~64, no.~8, pp. 871--874, 2003.

\bibitem{giguere_reattribution_2025}
\BIBentryALTinterwordspacing
S.~Giguère, M.~Beaudoin, L.~Dellazizzo, K.~Phraxayavong, S.~Potvin, and A.~Dumais, ``Reattribution of {Auditory} {Hallucinations} {Throughout} {Avatar} {Therapy}: {A} {Case} {Series},'' \emph{Reports}, vol.~8, no.~3, p. 113, Jul. 2025. [Online]. Available: \url{https://pmc.ncbi.nlm.nih.gov/articles/PMC12285938/}
\BIBentrySTDinterwordspacing

\bibitem{garety_digital_2024}
\BIBentryALTinterwordspacing
P.~A. Garety, C.~J. Edwards, H.~Jafari, R.~Emsley, M.~Huckvale, M.~Rus-Calafell, M.~Fornells-Ambrojo, A.~Gumley, G.~Haddock, S.~Bucci, H.~J. McLeod, J.~McDonnell, M.~Clancy, M.~Fitzsimmons, H.~Ball, A.~Montague, N.~Xanidis, A.~Hardy, T.~K.~J. Craig, and T.~Ward, ``\BIBforeignlanguage{en}{Digital {AVATAR} therapy for distressing voices in psychosis: the phase 2/3 {AVATAR2} trial},'' \emph{\BIBforeignlanguage{en}{Nature Medicine}}, vol.~30, no.~12, pp. 3658--3668, Dec. 2024, publisher: Nature Publishing Group. [Online]. Available: \url{https://www.nature.com/articles/s41591-024-03252-8}
\BIBentrySTDinterwordspacing

\bibitem{huckvale_avatar_2013}
M.~Huckvale, J.~Leff, and G.~Williams, ``Avatar {Therapy}: an audio-visual dialogue system for treating auditory hallucinations,'' in \emph{Interspeech}, 2013, pp. 392--396.

\bibitem{lee_sound_2022}
\BIBentryALTinterwordspacing
H.~Lee, R.~Jiang, Y.~Yoo, M.~Henry, and J.~R. Cooperstock, ``The {Sound} of {Hallucinations}: {Toward} a more convincing emulation of internalized voices,'' \emph{Conference on Human Factors in Computing Systems - Proceedings}, Apr. 2022, publisher: Association for Computing Machinery ISBN: 9781450391573. [Online]. Available: \url{https://dl.acm.org/doi/10.1145/3491102.3501871}
\BIBentrySTDinterwordspacing

\bibitem{Lee2022}
\BIBentryALTinterwordspacing
------, ``The sound of hallucinations: Toward a more convincing emulation of internalized voices,'' in \emph{Human Factors in Computing Systems (CHI)}. New Orleans, LA: ACM, Apr. 2022. [Online]. Available: \url{https://dl.acm.org/doi/10.1145/3491102.3501871}
\BIBentrySTDinterwordspacing

\bibitem{cooper_review_2024}
E.~Cooper, W.-C. Huang, Y.~Tsao, H.-M. Wang, T.~Toda, and J.~Yamagishi, ``A review on subjective and objective evaluation of synthetic speech,'' \emph{Acoustical Science and Technology}, vol.~45, no.~4, pp. 161--183, 2024.

\bibitem{huang_voicemos_2022}
\BIBentryALTinterwordspacing
W.~C. Huang, E.~Cooper, Y.~Tsao, H.-M. Wang, T.~Toda, and J.~Yamagishi, ``\BIBforeignlanguage{en}{The {VoiceMOS} {Challenge} 2022},'' in \emph{\BIBforeignlanguage{en}{Interspeech 2022}}. ISCA, Sep. 2022, pp. 4536--4540. [Online]. Available: \url{https://www.isca-archive.org/interspeech_2022/huang22f_interspeech.html}
\BIBentrySTDinterwordspacing

\bibitem{mittag_nisqa_2021}
\BIBentryALTinterwordspacing
G.~Mittag, B.~Naderi, A.~Chehadi, and S.~Möller, ``\BIBforeignlanguage{en}{{NISQA}: {A} {Deep} {CNN}-{Self}-{Attention} {Model} for {Multidimensional} {Speech} {Quality} {Prediction} with {Crowdsourced} {Datasets}},'' in \emph{\BIBforeignlanguage{en}{Interspeech 2021}}. ISCA, Aug. 2021, pp. 2127--2131. [Online]. Available: \url{https://www.isca-archive.org/interspeech_2021/mittag21_interspeech.html}
\BIBentrySTDinterwordspacing

\bibitem{Ghazanfar2008}
A.~A. Ghazanfar and D.~Rendall, ``Evolution of human vocal production,'' \emph{Current Biology}, vol.~18, no.~11, pp. R457--R460, 2008.

\bibitem{baumann_perceptual_2010}
O.~Baumann and P.~Belin, ``Perceptual scaling of voice identity: {Common} dimensions for different vowels and speakers,'' \emph{Psychological Research}, vol.~74, no.~1, pp. 110--120, 2010.

\bibitem{kreiman_information_2024}
J.~Kreiman, ``Information conveyed by voice quality,'' \emph{The Journal of the Acoustical Society of America}, vol. 155, no.~2, pp. 1264--1271, Feb. 2024, publisher: AIP Publishing.

\bibitem{latinus_norm-based_2013}
M.~Latinus, P.~McAleer, P.~E.~G. Bestelmeyer, and P.~Belin, ``Norm-based coding of voice identity in human auditory cortex,'' \emph{Current Biology}, vol.~23, no.~12, pp. 1075--1080, 2013.

\bibitem{orepic_bone_2023}
P.~Orepic, O.~A. Kannape, N.~Faivre, and O.~Blanke, ``Bone conduction facilitates self-other voice discrimination,'' \emph{Royal Society Open Science}, vol.~10, no.~2, Feb. 2023, publisher: The Royal Society.

\bibitem{WaveRNN}
N.~Kalchbrenner \emph{et~al.}, ``Efficient neural audio synthesis,'' in \emph{Proceedings of the 35th International Conference on Machine Learning (ICML)}, 2018.

\bibitem{HiFiGAN}
J.~Kong, J.~Kim, and J.~Bae, ``Hifi-gan: Generative adversarial networks for efficient and high fidelity speech synthesis,'' in \emph{Advances in Neural Information Processing Systems (NeurIPS)}, vol.~33, 2020.

\bibitem{Parselmouth}
Y.~Jadoul, B.~Thompson, and B.~de~Boer, ``Introducing parselmouth: A python interface to {Praat},'' \emph{Journal of Phonetics}, vol.~71, 2018.

\bibitem{Praat}
\BIBentryALTinterwordspacing
P.~Boersma and D.~Weenink, \emph{{Praat}: doing phonetics by computer}, 2024, computer program, Version 6.4.0. [Online]. Available: \url{http://www.praat.org/}
\BIBentrySTDinterwordspacing

\bibitem{Quatieri}
T.~F. Quatieri, \emph{Principles of Discrete-Time Speech Processing}. Upper Saddle River, NJ: Prentice Hall, 2001, explains F0 estimation and voiced/unvoiced classification standards.

\bibitem{harmonicity}
\BIBentryALTinterwordspacing
P.~Boersma and D.~Weenink, \emph{Praat Manual: Harmonicity}, 2026, accessed: 2026-04-03. [Online]. Available: \url{https://praat.org/manual/Harmonicity.html}
\BIBentrySTDinterwordspacing

\bibitem{ROC}
T.~Fawcett, ``An introduction to roc analysis,'' \emph{Pattern recognition letters}, vol.~27, no.~8, pp. 861--874, 2006.

\end{thebibliography}

\end{document}